\documentclass[12pt]{iopart}

\usepackage{graphicx}

\begin{document}
\title{Gauss-Bonnet brane-world cosmology without $Z_{2}$-symmetry}
\author{Kenichiro Konya}
\address{Institute for Cosmic Ray Research, 
University of Tokyo, Kashiwa 277-8582, Japan}
\ead{konya@icrr.u-tokyo.ac.jp}

\begin{abstract}
We consider a single 3-brane situated between two bulk spacetimes that possess the same
 cosmological constant, but whose metrics  do not possess a $Z_{2}$-symmetry. On each side of the brane, the bulk is a solution to Gauss-Bonnet gravity. This asymmetry  modifies junction conditions, and so new terms arise in the Friedmann equation. 
If $Z_2$-breaking terms become dominant, these behave cosmological constant at early times for some case, and might remove the initial singularity for other case. 
However, we show that these new terms  can not become dominant ones  under usual conditions when our brane is outside an event horizon .    
 We also show  that any brane-world scenarios of this type revert to a $Z_{2}$-symmetric form at late times, and hence rule out certain proposed scenarios.      
\end{abstract}


\maketitle

\section{Introduction}
There has been considerable interest in the brane-world scenario. Therein 
our universe is modeled by a 3-brane embedded in a higher-dimensional bulk spacetime. 
Of particular interest is Randall-Sundrum (RS) model, where a single brane is embedded 
in a five dimensional anti de Sitter (AdS$_{5}$) spacetime \cite{RS2}. Although the fifth 
dimension is noncompact, the graviton is localized at low energies on the brane due to 
the warped geometry of the bulk. \\
  ~~In four dimensions the Einstein tensor is the only second-rank tensor which satisfies the 
  following conditions \\
  $\bullet$ it is symmetric. \\
  $\bullet$ it has vanishing divergence.\\
  $\bullet$ it depends only on the metric and its first  two derivatives. \\
  $\bullet$ it is linear in the second derivatives of the metric.\\
  However, in D $>$ 4 dimensions, the Lovelock tensor satisfies similar conditions:  it is
  symmetric, divergence free, second order in the metric 
  but, quasi-linear in the second derivatives of the metric \cite{Lovelock,Lanczos}. 
  Thus, a natural extension to the RS model is to include such higher order curvature 
invariants in the bulk action. The Lovelock tensor arises from the variation of 
the Gauss-Bonnet (GB) term, 
  \begin{eqnarray}
  L_{GB} = R^{2} - 4R_{ab}R^{ab} + R_{abcd}R^{abcd}.
  \end{eqnarray}
 ~~String theory also provides us with a more compelling reason to include the GB term.
 The GB term appears as the next-to-leading order correction in the heterotic string effective 
 action, and it is ghost-free \cite{ghost}. Moreover, the graviton is localized in the GB 
 brane-world \cite{GBlocal} and deviations from Newton's law at low energies are less 
 pronounced than in the RS model \cite{GBnew}. This term is a topological invariant in 
 four dimensions, however in AdS$_{4}$ gravity the addition of this term has nontrivial 
 consequences \cite{GB4}. \\
 ~~Brane cosmologies with and without GB term have been investigated \cite{GBcosmo,asymcos,energy,asymcos3}. 
 Most brane-world scenarios assume a $Z_{2}$-symmetry about our brane. This is motivated 
 by a model derived from M-theory proposed by Horava and Witten \cite{HW}.  However, many
recent papers examine models that are not directly derived from M-theory: for example there 
 has been much interest in the one infinite extra dimension proposal. And there are at least 
 two ways in which the asymmetry might arise \cite{asym}. It is therefore interesting to analyze a 
 brane-world model without this symmetry \cite{asymcos,asymcos3,RSasymm,sp}.\\
 ~~The rest of this paper is organized as follows: in section 2 we present the complete setup we 
 consider; in section 3 we investigate the cosmological consequences without $Z_{2}$-symmetry; in section 4 some  conclusions are drawn.

\section{Einstein equations} 
We consider two 5D bulk spacetimes, $\mathcal{M}_{L}$ and $\mathcal{M}_{R}$, separated 
by a single 3-brane. $\mathcal{M}_{L,R}$ is a solution to Gauss-Bonnet gravity with a cosmological  
constant, $\Lambda <0$. This scenario is described by the following action,
 \begin{eqnarray}
S = S_{grav} + S_{brane},~~~~~~~~~~~~~~~~~~~\\
S_{grav} = \sum_{i=L,R}\frac{1}{2\kappa^{2}}\int_{\mathcal{M}_{i}}d^{5}x\sqrt{-g}\{
R-2\Lambda+\alpha L_{GB}\} ~~\nonumber\\ 
- \frac{1}{\kappa^{2}}\int_{\partial\mathcal{M}_{i}} d^{4}x\sqrt{-h}\left(K + 
2\alpha[J-2\hat{G}^{ab}K_{ab}] \right),\\
S_{brane} = \int_{brane}d^{4}x\sqrt{-h}~L_{brane}.~~~~~~~~~~~~~~~~~~~~~~~~~
\end{eqnarray}
 Here, $\kappa^{2}$ is the 5D gravitational constant and $\alpha > 0$ is a Gauss-Bonnet coupling.
 $h_{ab}$ is the induced 4D metric and is defined by 
 \begin{eqnarray}
h_{ab} = g_{ab} - n_{a}n_{b},
\end{eqnarray}
where $n^{a}$ is the spacetime unit vector, and it points away from the surface and into the 
adjacent space. The second term in $S_{grav}$  is a boundary term required for a well defined 
action principle \cite{boundary}. $\hat{G}_{ab}$ is the 4D Einstein tensor on the brane 
corresponding to $h_{ab}$. $K$ is the trace of the extrinsic curvature, defined by $K_{ab} = h^{c}_{~a}{}^B\nabla_{c}n_{b}$ where ${}^B\nabla^a$ is the covariant derivative associated 
with the bulk metric $g_{ab}$. 
$J$ is the trace of 
\begin{eqnarray}
J_{ab} = \frac{1}{3}(2KK_{ac}K^{c}_{~b} + K_{cd}K^{cd}K_{ab} 
-2K_{ac}K^{cd}K_{db}-K^{2}K_{ab}).
\end{eqnarray}

The variation of the action $S$ gives 
\begin{eqnarray}
G_{ab} + 2\alpha H_{ab} + \Lambda g_{ab} = \kappa^{2}S_{ab}\delta(\partial M),
\end{eqnarray}
where
\begin{eqnarray}
G_{ab} = R_{ab} - \frac{1}{2}g_{ab}R,~~~~~~~~~~~~~~~~~~~~~\\
H_{ab} = RR_{ab} -2R_{ac}R^{c}_{~b} -2R^{cd}R_{acbd}
+R_{a}^{~cde}R_{bcde} -\frac{1}{4}g_{ab}L_{GB},~~~~~~
\end{eqnarray}
and
\begin{eqnarray}
S_{ab} = -2\frac{\delta L_{brane}}{\delta h^{ab}} + h_{ab}L_{brane}.
\end{eqnarray}
 ~~Conservation of energy momentum of the matter on the brane follows from the Gauss-Codazzi 
 equations.  Thus we have \cite{energy}
 \begin{eqnarray}
\nabla^{a}S_{ab} = 0,
\end{eqnarray}
where $\nabla^a$ is the covariant derivative associated with the brane metric $h_{ab}$

\subsection{The bulk}
We assume that our brane is homogeneous and isotropic. 
Then, from the generalized Birkhoff's theorem, the bulk metric can be written in the following
 form \cite{sp}:
\begin{eqnarray}
ds^{2} = -f_{L,R}(a)dt^{2} + \frac{da^{2}}{f_{L,R}(a)} +a^{2}\Omega_{ij}dx^{i}dx^{j},
\end{eqnarray}
where $\Omega_{ij}$ is the three dimensional metric of space with constant curvature $k = 
-1,0,1$. For the time being, let us drop the index $L,R$, as the following analysis will apply on 
both sides of the brane. The solution of the field equation $G_{ab} + 2\alpha H_{ab} + \Lambda g_{ab} = 0$ is 
given by 
\begin{eqnarray}
f(a) = k + \frac{a^{2}}{4\alpha}\left(1\mp\sqrt{1+\frac{4}{3}\alpha\Lambda+8\alpha\frac{\mu}{a^{4}}}
\right),
\end{eqnarray}
 with $\mu$ being an arbitrary constant. Other solutions do exist for special values of 
 $k,~\Lambda$, and $\alpha$ \cite{sp}, but we will not consider them here. The constant $\mu$ 
 is related to the black hole mass by the relation
 \begin{eqnarray}
M_{BH} = \frac{3V\mu}{\kappa^{2}},
\end{eqnarray}
where $V$ is the volume of the 3D space \cite{BH}. 
In order to avoid a classical instability \cite{naked}, we must have $\mu \geq 0$.\\
~~From Eq. (13) we find that the metric has a singularity at $a =0$. For the ($-$) branch, this singularity is covered by an event horizon if $k\leq0$ or $k=1$ and $\mu\geq2\alpha$. For such cases, 
 the event horizon ($a = a_{h}$) is 
\begin{eqnarray}
a^{2}_{h} = \frac{3k}{\Lambda} + \sqrt{\frac{9k^{2}+12k^{2}\alpha\Lambda-6\mu\Lambda}{\Lambda^{2}}}.
\end{eqnarray}
This is not the case for the (+) branch. So, to shield this naked singularity we must cut the spacetime 
off at some small value of $a$ for the ($-$) branch with $k=1$ and $\mu<2\alpha$, and for the (+) branch. 
This can be done by introducing a second brane at $a\sim M^{-1}_{cut}$ \cite{shield}.

\subsection{The brane}
We define the position of the brane as $a=a(\tau)$ and $t=t(\tau)$ which is parameterized by the 
proper time on the brane. Then the induced metric of the 3-brane is 
\begin{eqnarray}
ds^{2} = -d\tau^{2} + a(\tau)^{2}\Omega_{ij}dx^{i}dx^{j}.
\end{eqnarray}
 The tangent vector of the brane is $u^{a}_{L,R} = (\dot{t}_{L,R},0,0,0,\dot{a})$, where the dot denotes 
 the derivative with respect to a proper time $\tau$. 
 The normal vector is $n_{La} = (-\dot{a},0,0,0,\dot{t}_{L})$, $n_{Ra}= (\dot{a},0,0,0,-\dot{t}_{R})$. Normalization of $n^{a}$ implies 
 \begin{eqnarray}
-f_{L,R}^{2}\dot{t}_{L,R}^{2} + \dot{a}^{2} = -f_{L,R}.
\end{eqnarray}
And the requirement that the metrics must be continuous at the brane  implies \cite{asymcos3}
\begin{eqnarray}
t_{L} = \left. \frac{t_Rf_R}{f_L}\sqrt{\frac{\dot{a}^2+f_L}{\dot{a}^2+f_R}}\right|_{brane}.
\end{eqnarray}
~~From Eq. (7) we obtain  the generalized Israel's junction condition \cite{energy,gIsrael}
\begin{eqnarray}
[K_{\mu\nu}]_{-} -h_{\mu\nu}[K]_{-} + 2\alpha(3[J_{\mu\nu}]_{-}-h_{\mu\nu}[J]_{-} 
-2P_{\mu\rho\nu\sigma}[K^{\rho\sigma}]_{-}) = -\kappa^{2}S_{\mu\nu},
\end{eqnarray}
where 
\begin{eqnarray}
P_{\mu\nu\rho\sigma} = R_{\mu\nu\rho\sigma} +2h_{\mu[\sigma}R_{\rho]\nu}
+ 2h_{\nu[\rho}R_{\sigma]\mu} + Rh_{\mu[\rho}h_{\sigma]\nu} 
\end{eqnarray} 
is  the divergence free part of the Riemann tensor. We have introduced $[X]_{-} \equiv X_{R} - X_{L}$.\\
~~We take the brane matter to be a perfect fluid, so $S_{ab} = (\rho + p)u_{a}u_{b} + ph_{ab}$. 
The $(\tau,\tau)$ component of Eq. (19) is then 
\begin{eqnarray}
\frac{3f_{R}\dot{t}_{R}}{a} + \frac{3f_{L}\dot{t}_{L}}{a} + 2\alpha\left[
-\frac{2f_{R}^{3}\dot{t}_{R}^{3}}{a^{3}}-\frac{2f_{L}^{3}\dot{t}_{L}^{3}}{a^{3}} 
+6(\dot{a}^{2}+k)\left(\frac{f_{R}\dot{t}_{R}}{a^{3}} + \frac{f_{L}\dot{t}_{L}}{a^{3}}\right)\right] \nonumber \\
= \kappa^{2}\rho.
\end{eqnarray}
From Eqs. (13), (17) and (21) we have a cubic equation for $H^{2}$ 
\begin{eqnarray}
A^{3} - \frac{9\left(b_{L}^{2/3}-b_{R}^{2/3}\right)^{2}}{256\alpha^{2} \kappa^{4}\rho^{2}} A^{2} 
\nonumber \\
+ \left[-\frac{3\left(b_{L}^{2/3}+b_{R}^{2/3}\right)}{128\alpha^{2}} \mp \frac{3\left(b_{L}^{2/3} - b_{R}^{2/3}\right)\left(b_{L}-b_{R}\right)}{512\alpha^{3} \kappa^{4}\rho^{2}}\right]A \nonumber \\
+\left[-\frac{ \kappa^{4}\rho^{2}}{256\alpha^{2}}\mp \frac{b_{L}+b_{R}}{512\alpha^{3}} - \frac{\left(b_{L}-b_{R}\right)^{2}}{4096\alpha^{4}\kappa^{4}\rho^{2}}\right] = 0,
\end{eqnarray}
where
\begin{eqnarray}
A = H^{2}+\frac{k}{a^{2}} + \frac{1}{4\alpha},\\ 
b_{L,R}^{1/3}= \left(1+\frac{4}{3}\alpha\Lambda+\frac{8\alpha\mu_{L,R}}{a^{4}}\right)^{1/2}.
\end{eqnarray}
The $\mp$ signs in Eqs. (22) correspond to those in Eq. (13).

\section{The effect of no Z$_2$-symmetry}
In this section we study the cosmological effects of no $Z_{2}$-symmetry. Here, the GB term may be considered as the lowest-order stringy correction to the 5D 
Einstein action. In this case, $\alpha|R^{2}|< |R|$, so that 
\begin{eqnarray}
\alpha < l^{2}
\end{eqnarray}
 where $l$ is the bulk curvature scale, $|R|\sim l^{-2}= \left(1\mp\sqrt{1+4\alpha\Lambda/3}\right)/ 4\alpha$. For the ($-$) branch this reduces to 
 the RS relation $l^{-2} = -\Lambda/6$ when $\alpha = 0$. Note that there is an 
 upper limit to the GB coupling: 
 \begin{eqnarray}
\alpha < \frac{l^2}{4} ~~~~~~~~~~(-)~\textmd{branch}, \\
\frac{l^2}{4} < \alpha < \frac{l^2}{2}~~~\textmd{(+)~branch}.
\end{eqnarray}
These conditions are consistent with Eq. (25).

\subsection{The ($-$) branch}
\begin{figure}[t]
\begin{center}
\includegraphics[width=8cm]{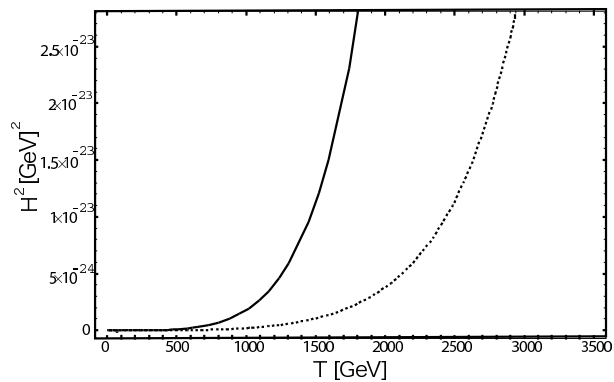}
\caption{Illustrating the relationship between $H^{2}$ and $T$. The solid line denotes the $Z_{2}$-braking case, and the doted line denotes the $Z_{2}$-symmetric one. The adopted parameters are
$\alpha = 10^{20}$ (GeV)$^{-2}$, $\Lambda = -2.4\times 10^{-23}$ (GeV)$^{2}$, $\gamma = 
2.9\times10^{-51}\times\left(g_{*}/10.35\right)$ (GeV)$^{4}$, and $C^{2} = 10^{-2}$.}
\label{-branch}
\end{center}
\end{figure}

We first consider the ($-$) branch and we assume $k=0$ for simplicity. 
The Friedmann equation is shown in Appendix A.1. 
We find that there are more $Z_2$-breaking terms in the GB case than in the RS case.   
This is of course because including the GB term  gives rise to extra terms in  eq. (7). 
From Eqs. (15) and (A.9) we find that the third term in Eq. (24) is always smaller than the other terms. 
And then the $Z_2$-breaking terms outside the square root of Eq. (A.2) approximately 
scale as $1/(\rho^2 a^8)^n$ where  $n = 1,2,3$ and  the $Z_2$-breaking terms in the square
 root of Eq. (A.2) approximately scale as $1/(\rho^2 a^8)^n$ where  $n = 1,2,3,4$.
Therefore, when $\sigma > \lambda$ (here, we take the usual assumption that 
$\rho = \sigma + \lambda$ where $\sigma$ is the ordinary matter energy density and $\lambda$  
is  the brane tension), the $Z_2$-breaking terms behave like positive cosmological constant term
for radiation dominant universe $(\sigma = \gamma/a^{4})$. And so we might have an inflation without 
any other fields in the early universe. 
At late times ($\sigma < \lambda$)  the $Z_2$-breaking terms rapidly  decrease. Thus, the   effects of 
$Z_{2}$-breaking terms are no longer significant and we obtain the standard cosmology. \\
~~In order to obtain the $Z_2$-breaking term dominant regime,  one of the $Z_2$-breaking terms 
must be larger than all the $Z_2$-symmetry terms during $\sigma > \lambda$. From this requirement 
we have $C^{2} = \alpha(\mu_{L}-\mu_{R})^{2}/\kappa^{4}\gamma^{2}b^{1/3} > 1/18$. \\
 ~~Here, the third term in the right hand side of the Eq. (A.6) scales just like radiation. Hence it is called 
 the dark radiation. Dark radiation affects both big bang nucleosynthesis and the cosmic microwave 
 background. Accordingly such observations constraint on $\mu$ \cite{darkradiatiion}:
 \begin{eqnarray}
\frac{\mu_{L}+\mu_{R}}{2b^{1/3}a^{4}} < \frac{\pi^{2}}{30}\frac{\Delta g_{*}T_{N}^{4}}{3M_{pl}^{2}},
\end{eqnarray}
where $\Delta g_{*}$ is the deviation from number of effective relativistic degrees of freedom ($g_{*}$), 
 and $T_{N}$ is the temperature of nucleosynthesis. 
At the time of  nucleosynthesis, the energy density is 
\begin{eqnarray}
\sigma(t_{N}) = \frac{\gamma}{a^{4}(t_{N})}=\frac{\pi^{2}g_{*}T_{N}^{4}}{30}.
\end{eqnarray}
From Eqs. (28) and (29)
\begin{eqnarray}
\frac{\mu_{L}+\mu_{R}}{\gamma} < \frac{2b^{1/3}\Delta g_{*}}{3M_{pl}^{2}g_{*}}\sim \frac{1}{10M_{pl}^{2}},
\end{eqnarray}
where we have taken the standard values $g_{*}=10.75$ and $\Delta g< 2$.
 Using Eqs. (30), (A.7), (A.8) and (A.9) we find that $C^{2}$ is bounded above:
 \begin{eqnarray}
C^{2} \hspace{0.3em}\raisebox{0.4ex}{$<$}\hspace{-0.75em}\raisebox{-.7ex}{$\sim$}\hspace{0.3em} \frac{\alpha\mu^{2}}{b^{1/3}\kappa^{4}\gamma^{2}}
\sim \frac{1-b^{1/3}}{400b} \hspace{0.3em}\raisebox{0.4ex}{$<$}\hspace{-0.75em}\raisebox{-.7ex}{$\sim$}\hspace{0.3em} 8\times10^{-4}.
\end{eqnarray}
Note that $C^{2} > 1/18$ is needed for the $Z_{2}$-breaking term dominant regime.
Thus, we find that  the $Z_{2}$-symmetric term is always dominant in general.  And we have the usual evolution of the universe (see figure \ref{-branch}) .\\
  
\subsection{The (+) branch} 
\begin{figure}[t]
\begin{center}
\includegraphics[width=8cm]{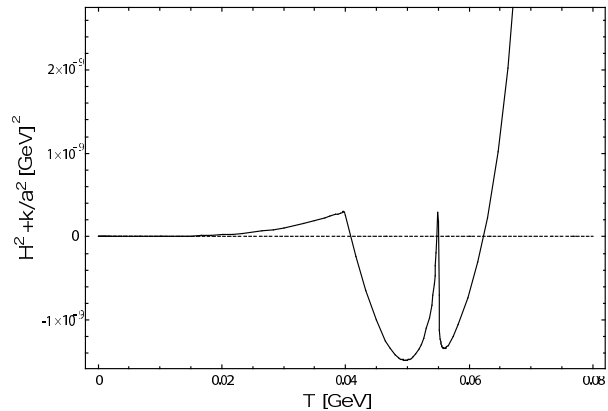}
\caption{Illustrating the relationship between $H^{2}+k/a^{2}$ and $T$. The solid line denotes the $Z_{2}$-braking case, and the doted line denotes the $Z_{2}$-symmetric one. 
The adopted parameters are $\alpha = 10^{8}$ (GeV)$^{-2}$, $\Lambda = -1.0\times 10^{-11}$ (GeV)$^{2}$, $\gamma = 
2.9\times10^{-51}\times\left(g_{*}/10.35\right)$ (GeV)$^{4}$, and $\mu = 7.8\times10^{-55}$ (GeV)$^2$.}
\label{+branch}
\end{center}
\end{figure}

In this section we consider the (+) branch. It was claimed that this branch solution is classically 
unstable to small perturbations and yielding a graviton ghost \cite{unstable,sp}. However, it turns out that this branch solution is classically stable to small perturbations due to the positive Abbott-Deser (AD) energy \cite{stable}. \\
~~ In this branch, the usual assumption that the GB 
energy scale is greater than the RS energy scale  is inconsistent with Eq. (27) (See the Appendix A.2).  
So, we investigate the case that the GB energy scale is smaller than the RS energy scale. \\
$\bullet$Case 1. $\left(m_{RS}^{8}  \hspace{0.3em}\raisebox{0.4ex}{$>$}\hspace{-0.75em}\raisebox{-.7ex}{$\sim$}\hspace{0.3em}\ 2m_{GB}^{8}\right)$ \\
~~The Friedmann equation is the one where we have changed the sign of $b_{L,R}^{1/3}$ in the $(-)$ 
branch and is shown in the Appendix  A.2. Note that there is not a dark radiation term in Eq (A.13).  
Therefore, we will be able to have the $Z_{2}$-breaking term dominant universe.  We do not have an event horizon in this branch. Thus, the third term in Eq. (24) can be larger than the other terms not 
as in the $(-)$ branch. Then the $Z_2$-breaking terms outside the square root of Eq. (A.11) 
approximately scale as 
$(-1)^{n+1}/(\rho^{2n} a^{6+6n})$ where $n = 1,2,3$ and the $Z_2$-breaking terms in the square root 
of Eq. (A.11) approximately scale as 
$(-1)^{n+1}/(\rho^{2n} a^{6+6n})$ where $n = 1,2,3,4.$ 
 Therefore,  the Z$_2$-breaking term will  
 behave like positive cosmological constant term for radiation 
 dominant universe. At late times the Z$_2$-breaking terms rapidly damp and standard cosmology 
 is recovered. \\
 ~~Here, $H^2 + k/a^2$ can be negative if the dark radiation term is so large (see figure \ref{+branch}). 
 This is because we assume that the RS energy scale is larger than the GB one and 
  If $k\geq0$, this is inconsistent with the positivity of $H^{2}+k/a^{2}$. And it follows that $a$ is 
  bounded from below as $a^{6} \geq 8\mu\sqrt{2\alpha\mu}/\kappa^{4}\lambda^{2}.$ 
  Therefore, we can avoid an initial singularity without introducing a second brane. 
For the case of $k<0$, if $\mu\hspace{0.3em}\raisebox{0.4ex}{$>$}\hspace{-0.75em}\raisebox{-.7ex}{$\sim$}\hspace{0.3em} 300\alpha,$ $a$ is bounded 
from below as the case of $k\geq0$.  In such cases we cannot have an inflation due to 
the $Z_2$-breaking.
If $\mu<300\alpha$, $a$ is not bounded from below and we have an inflation and unusual evolution of 
universe. \\
$\bullet$Case 2. $\left(  1\hspace{0.3em}\raisebox{0.4ex}{$<$}\hspace{-0.75em}\raisebox{-.7ex}{$\sim$}\hspace{0.3em}m_{RS}^{8}/m_{GB}^{8}\hspace{0.3em}\raisebox{0.4ex}{$<$}\hspace{-0.75em}\raisebox{-.7ex}{$\sim$}\hspace{0.3em}2\right)$  \\
~~From the Friedmann equation (A.18)  we find that there is a dark radiation term and that Newton's constant evolves.  The evolution of Newton constant has an effect on various theories such as 
cosmology, astrophysics, geophysics, etc. Thus $\mu$ is restricted from such observations (see 
\cite{Gravity}, reference therein) in addition to the restriction for the dark radiation term. From these 
restrictions we find that the $Z_{2}$-breaking terms cannot become dominant ones.

 \section{Conclusions}
In this paper we have considered the Gauss-Bonnet brane-world cosmology without 
$Z_{2}$-symmetry. Relaxing the $Z_{2}$-symmetry gives more extra terms than RS case. \\
~~For the ($-$) branch, these $Z_{2}$-breaking terms can not become dominant outside the event horizon in general. 
If these term can be dominant,  these behave cosmological constant at early times. \\
~~For the (+) branch, if $m_{RS}<m_{GB}$, the late time  cosmology is not compatible with the 
 standard one. So, we have considered the $m_{RS}>m_{GB}$ case. 
In this case, the late time cosmology is compatible with the standard one. 
 In the case of $m_{RS}^{8}>2m_{GB}^{8}$,
 we have shown that initial singularity problem might be avoided by the $Z_{2}$-breaking terms.
 In the case of $1<m_{RS}^8/m_{GB}^8<2$, we have shown that the Newton constant evolves  
 and  that we can not  have the $Z_{2}$-breaking term dominant universe because of the restriction  
 for the dark radiation and Newton constant. \\
 ~~We also note that the effects of the $Z_{2}$-breaking terms decrease with time like RS case 
 \cite{RSasymm}. Therefore the scenario without this symmetry at late times are not viable.

\ack
 We would like to thank M.~Kawasaki and T.~Takahashi for the helpful advices.

\appendix
\section{Friedmann equation}
\subsection{The $(-)$ brach}
We first consider the $(-)$ branch. The single real solution of the cubic equation  (22)  is 
\begin{eqnarray}
H^{2} + \frac{k}{a^{2}} + \frac{1}{4\alpha} - \frac{3\left(b_{L}^{2/3}-b_{R}^{2/3}\right)^{2}}{2^{8}\alpha^{2} \kappa^{4}\rho^{2}} = c_{+} + c_{-},
\end{eqnarray}
where
\begin{eqnarray}
 c_{\pm} = \frac{1}{2^{8}}\left\{\frac{2^{15} \kappa^{4}\rho^{2}}{\alpha^{2}}+\frac{2^{14}\left(b_{L}+b_{R}\right)}{\alpha^{3}}  \right.
+\frac{2^{8}}{\alpha^{4} \kappa^{4}\rho^{2}}\left(b_{L}^{1/3}- b_{R}^{1/3}\right)^{2} \nonumber \\ 
\times  \left(17b_{L}^{4/3}
+34b_{L}b_{R}^{1/3}+42b_{L}^{2/3}b_{R}^{2/3} +34b_{L}^{1/3}b_{R} +17b_{L}^{4/3}\right) \nonumber\\ 
+\frac{2^{6}\cdot3^{2}}{\alpha^{5} \kappa^{8}\rho^{4}}\left(b_{L}^{1/3}-b_{R}^{1/3}\right)^{4}\left(b_{L}^{1/3}+b_{R}^{1/3}\right)^{3}  \left(b_{L}^{2/3} +b_{L}^{1/3}b_{R}^{1/3}+b_{R}^{2/3}\right) \nonumber \\ 
+ \frac{3^{3}\left(b_{L}^{2/3}-b_{R}^{2/3}\right)^{6}}{\alpha^{6} \kappa^{12}\rho^{6}} \nonumber \\
\pm 2^{8}\kappa^{2}\rho\left[\frac{ 2^{14}\kappa^{4}\rho^{2}}{\alpha^{4}}+\frac{2^{14}(b_{L}+b_{R})}{\alpha^{5}}  \right. 
+\frac{2^{8}}{\alpha^{6}\kappa^{4}\rho^{2}} \left(b_{L}^{1/3}-b_{R}^{1/3}\right)^{2} \nonumber \\
\times\left(25b_{L}^{4/3}+50b_{L}b_{R}^{1/3}+42b_{L}^{2/3}b_{R}^{2/3} \right. 
+ \left. 50b_{L}^{1/3}b_{R}+25b_{R}^{4/3}\right) \nonumber \\ 
+\frac{2^{6}}{\alpha^{7} \kappa^{8}\rho^{4}}\left(b_{L}^{1/3}-b_{R}^{1/3}\right)^{4}\left(b_{L}^{1/3}+b_{R}^{1/3}\right)  \left(19b_{L}^{4/3}
+57b_{L}b_{R}^{1/3} +64b_{L}^{2/3}b_{R}^{2/3} \right. \nonumber \\ \left.
+57b_{L}^{1/3}b_{R}+19b_{R}^{4/3}\right) \nonumber \\
+\frac{2^{2}}{\alpha^{8} \kappa^{12}\rho^{6}}\left(b_{L}^{1/3}-b_{R}^{1/3}\right)^{6}\left(2b_{L}^{1/3}+b_{R}^{1/3}\right)\left(b_{L}^{1/3}+2b_{R}^{1/3}\right) \nonumber \\
\times\left(14b_{L}^{4/3}+49b_{L}b_{R}^{1/3}+66b_{L}^{2/3}b_{R}^{2/3}+49b_{L}^{1/3}b_{R}+14b_{R}^{4/3}\right)+  \nonumber \\ \left.\left. 
\frac{1}{\alpha^{9} \kappa^{16}\rho^{8}}\left(b_{L}^{1/3}-b_{R}^{1/3}\right)^{8}\left(b_{L}^{1/3}+b_{R}^{1/3}\right)^{3}\left(2b_{L}^{2/3}+5b_{L}^{1/3}b_{R}^{1/3}+  2b_{R}^{2/3}\right)^{2}
\right]^{1/2}
\right\}^{1/3}.
\end{eqnarray}
Eqs. (11), (A.1) and (A.2)  are sufficient to determine the cosmic dynamics on the brane if an equation of 
state is specified for the matter source. Such an analysis can be simplified by another form of solution:
\begin{eqnarray}
H^{2} + \frac{k}{a^{2}} + \frac{1}{4\alpha} - \frac{3\left(b_{L}^{2/3}-b_{R}^{2/3}\right)^{2}}{2^{8}\alpha^{2} \kappa^{4}\rho^{2}} = 2q\cosh\frac{2}{3}x,
\end{eqnarray}
where 
\begin{eqnarray}
q = \frac{1}{2^{8}}\left[\frac{2^{9}\left(b_{L}^{2/3}+b_{R}^{2/3}\right)}{\alpha^{2}} +\frac{2^{7}}{\alpha^{3} \kappa^{4}\rho^{2}}\left(b_{L}^{1/3}-b_{R}^{1/3}\right)^{2} \right. \nonumber \\
\times\left(b_{L}+2b_{L}^{2/3}b_{R}^{1/3}+2b_{L}^{1/3}b_{R}^{2/3}+b_{R}\right) 
\left. +\frac{9\left(b_{L}^{2/3}-b_{R}^{2/3}\right)^{4}}{\alpha^{4}\kappa^{8}\rho^{4}}\right]^{1/2}, 
\end{eqnarray}
\begin{eqnarray}
\cosh2x =  \left[2^{15}\alpha^{4} \kappa^{16}\rho^{8}+ 2^{14}\left(b_{L}+b_{R}\right)\alpha^{3} \kappa^{12}\rho^{6}+2^{8}\left(b_{L}^{1/3}-b_{R}^{1/3}\right)^{2} \right. \nonumber \\
\times\left(17b_{L}^{4/3}+34b_{L}b_{R}^{1/3}+42b_{L}^{2/3}b_{R}^{2/3}  +34b_{L}^{1/3}b_{R} +17b_{L}^{4/3}\right)\alpha^{2} \kappa^{8}\rho^{4} \nonumber \\
+2^{6}\cdot9\left(b_{L}^{1/3}-b_{R}^{1/3}\right)^{4}\left(b_{L}^{1/3}+b_{R}^{1/3}\right)^{3}\left(b_{L}^{2/3} +b_{L}^{1/3}b_{R}^{1/3}+b_{R}^{2/3}\right)\alpha \kappa^{4}\rho^{2}~~~~~~~~\nonumber \\
\left. +27\left(b_{L}^{2/3}-b_{R}^{2/3}\right)^{6} \right] / (2^{24}\alpha^{6}\kappa^{12}\rho^{6}q^{3}).
\end{eqnarray}
These solutions reduce to the GB case with $Z_{2}$-symmetry when $b_{L}=b_{R}$. So the absence 
of the $Z_{2}$-symmetry gives rise to above extra terms in the Friedmann equation. And  we find that 
the effects of the $Z_{2}$-breaking terms decrease with increasing time. This can be seen  in RS case \cite{RSasymm}. Therefore the scenario 
without this symmetry at late tine (\cite{woZ2}) are not viable even in the GB case.\\ 
~~The standard form of the Friedmann equation must be recovered at low energy. 
 Setting  $\kappa^{4}\alpha\rho^{2}$ and 
$\alpha\mu/a^{4}$ as small variables, the Friedmann equation is approximated as 
\begin{eqnarray}
H^{2} +\frac{k}{a^{2}}  = \frac{b^{1/3}- 1}{4\alpha} +\frac{\kappa^{4}(\sigma+\lambda)^{2}}{36b^{2/3}} 
+ \frac{\mu_{L}+\mu_{R}}{2b^{1/3}a^{4}} + \frac{9(\mu_{L}-\mu_{R})^{2}}{4\kappa^{4}(\sigma+\lambda)^{2}a^{8}},  
\end{eqnarray}
 where $b^{1/3}= \sqrt{1+4\alpha\Lambda/3}.$  We  assume that the GB 
energy scale( $m_{GB}= (2b/\kappa^{4}\alpha)^{1/8}$)   is greater than the RS energy scale 
($m_{RS} = \lambda^{1/4}$) \cite{GBRS}. This assumption comes from the consideration that 
the GB term is a correction to the RS gravity. \\
~~Eq. (A.6) reduces to the RS case without $Z_{2}$-symmetry 
when $\alpha=0$ \cite{RSasymm}.
In order to obtain standard cosmology at late times we need to make the identification,
\begin{eqnarray}
\kappa_{4}^{2} = \frac{1}{M_{pl} ^{2}} = \frac{\kappa^{4}\lambda}{6b^{2/3}} ,
\end{eqnarray}
where $M_{pl}$ is the 4D reduced Planck mass. 4D cosmological constant vanishes when the tension satisfies 
\begin{eqnarray}
\lambda = \frac{3}{2}\left(1-b^{1/3}\right)\frac{1}{\alpha\kappa_{4}^{2}}
\end{eqnarray}
 By Eqs. (A.7) and (A.8), the requirement ($m_{GB}>m_{RS}$) become 
\begin{eqnarray}
\alpha < \frac{l^{2}}{22}.
\end{eqnarray}
This condition is consistent with Eq.(26).

\subsection{The $(+)$ branch}
\begin{figure}[t]
\begin{center}
\includegraphics[width=8cm]{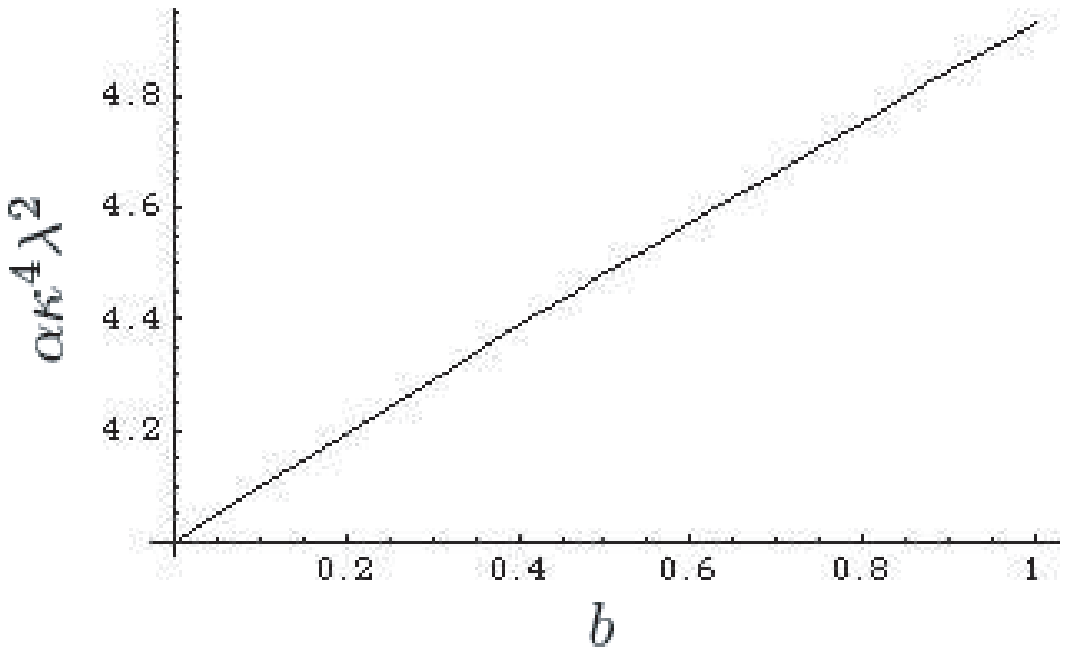}
\caption{Illustrating the relationship among the $\alpha\kappa^{4}\lambda^{2}$ and $b$.
 Note that $m_{RS}^{8}=\lambda^{2}$, $m_{GB}^{8}= 2b/\kappa^{4}\alpha$.}
\label{lambda}
\end{center}
\end{figure}

Next, we consider the $(+)$ branch. The solution of the cubic equation (21) is the one where 
 we have changed the sign of $b^{1/3}_{L,R}$ in the ($-$) branch. Thus,  the effects of the 
 $Z_{2}$-breaking terms also decrease with increasing time in this case.  The Friedmann equation 
 now becomes  
\begin{eqnarray}
H^{2} + \frac{k}{a^{2}} + \frac{1}{4\alpha} - \frac{3\left(b_{L}^{2/3}-b_{R}^{2/3}\right)^{2}}{2^{8}\alpha^{2} \kappa^{4}\rho^{2}} = c_{+} + c_{-},
\end{eqnarray}
where
\begin{eqnarray}
 c_{\pm} = \frac{1}{2^{8}}\left\{\frac{2^{15} \kappa^{4}\rho^{2}}{\alpha^{2}}-\frac{2^{14}\left(b_{L}+b_{R}\right)}{\alpha^{3}}  \right.
+\frac{2^{8}}{\alpha^{4} \kappa^{4}\rho^{2}}\left(b_{L}^{1/3}- b_{R}^{1/3}\right)^{2} \nonumber \\ 
\times  \left(17b_{L}^{4/3}
+34b_{L}b_{R}^{1/3}+42b_{L}^{2/3}b_{R}^{2/3} +34b_{L}^{1/3}b_{R} +17b_{L}^{4/3}\right) \nonumber\\ 
-\frac{2^{6}\cdot3^{2}}{\alpha^{5} \kappa^{8}\rho^{4}}\left(b_{L}^{1/3}-b_{R}^{1/3}\right)^{4}\left(b_{L}^{1/3}+b_{R}^{1/3}\right)^{3}  \left(b_{L}^{2/3} +b_{L}^{1/3}b_{R}^{1/3}+b_{R}^{2/3}\right) \nonumber \\ 
+ \frac{3^{3}\left(b_{L}^{2/3}-b_{R}^{2/3}\right)^{6}}{\alpha^{6} \kappa^{12}\rho^{6}} \nonumber \\
\pm 2^{8}\kappa^{2}\rho\left[\frac{ 2^{14}\kappa^{4}\rho^{2}}{\alpha^{4}}-\frac{2^{14}(b_{L}+b_{R})}{\alpha^{5}}  \right. 
+\frac{2^{8}}{\alpha^{6}\kappa^{4}\rho^{2}} \left(b_{L}^{1/3}-b_{R}^{1/3}\right)^{2} \nonumber \\
\times\left(25b_{L}^{4/3}+50b_{L}b_{R}^{1/3}+42b_{L}^{2/3}b_{R}^{2/3} \right. 
+ \left. 50b_{L}^{1/3}b_{R}+25b_{R}^{4/3}\right) \nonumber \\ 
-\frac{2^{6}}{\alpha^{7} \kappa^{8}\rho^{4}}\left(b_{L}^{1/3}-b_{R}^{1/3}\right)^{4}\left(b_{L}^{1/3}+b_{R}^{1/3}\right)  \left(19b_{L}^{4/3}
+57b_{L}b_{R}^{1/3} +64b_{L}^{2/3}b_{R}^{2/3} \right. \nonumber \\ \left.
+57b_{L}^{1/3}b_{R}+19b_{R}^{4/3}\right) \nonumber \\
+\frac{2^{2}}{\alpha^{8} \kappa^{12}\rho^{6}}\left(b_{L}^{1/3}-b_{R}^{1/3}\right)^{6}\left(2b_{L}^{1/3}+b_{R}^{1/3}\right)\left(b_{L}^{1/3}+2b_{R}^{1/3}\right) \nonumber \\
\times\left(14b_{L}^{4/3}+49b_{L}b_{R}^{1/3}+66b_{L}^{2/3}b_{R}^{2/3}+49b_{L}^{1/3}b_{R}+14b_{R}^{4/3}\right)-\frac{1}{\alpha^{9} \kappa^{16}\rho^{8}}  \nonumber \\ \left.\left. 
\times\left(b_{L}^{1/3}-b_{R}^{1/3}\right)^{8}\left(b_{L}^{1/3}+b_{R}^{1/3}\right)^{3}\left(2b_{L}^{2/3}+5b_{L}^{1/3}b_{R}^{1/3}+  2b_{R}^{2/3}\right)^{2}
\right]^{1/2}
\right\}^{1/3}.
\end{eqnarray}
Another form of solution is easy to handle:
\begin{eqnarray}
H^{2} + \frac{k}{a^{2}} + \frac{1}{4\alpha} - \frac{3\left(b_{L}^{2/3}-b_{R}^{2/3}\right)^{2}}{2^{8}\alpha^{2} \kappa^{4}\rho^{2}} = 2q\cosh\frac{2}{3}x,
\end{eqnarray}
where 
\begin{eqnarray}
q = \frac{1}{2^{8}}\left[\frac{2^{9}\left(b_{L}^{2/3}+b_{R}^{2/3}\right)}{\alpha^{2}} -\frac{2^{7}}{\alpha^{3} \kappa^{4}\rho^{2}}\left(b_{L}^{1/3}-b_{R}^{1/3}\right)^{2} \right. \nonumber \\
\times\left(b_{L}+2b_{L}^{2/3}b_{R}^{1/3}+2b_{L}^{1/3}b_{R}^{2/3}+b_{R}\right) 
\left. +\frac{9\left(b_{L}^{2/3}-b_{R}^{2/3}\right)^{4}}{\alpha^{4}\kappa^{8}\rho^{4}}\right]^{1/2}, 
\nonumber \\ 
\end{eqnarray}
\begin{eqnarray}
\cosh2x =  \left[2^{15}\alpha^{4} \kappa^{16}\rho^{8}- 2^{14}\left(b_{L}+b_{R}\right)\alpha^{3} \kappa^{12}\rho^{6} 
+2^{8}\left(b_{L}^{1/3}-b_{R}^{1/3}\right)^{2}  \right.     \nonumber \\
\times \left(17b_{L}^{4/3}+34b_{L}b_{R}^{1/3}+42b_{L}^{2/3}b_{R}^{2/3} +34b_{L}^{1/3}b_{R} +17b_{L}^{4/3}\right)\alpha^{2} \kappa^{8}\rho^{4}    \nonumber \\
-2^{6}\cdot9\left(b_{L}^{1/3}-b_{R}^{1/3}\right)^{4} \left(b_{L}^{1/3}+b_{R}^{1/3}\right)^{3}\left(b_{L}^{2/3} +b_{L}^{1/3}b_{R}^{1/3}+b_{R}^{2/3}\right)\alpha \kappa^{4}\rho^{2} \nonumber \\
\left. +27\left(b_{L}^{2/3}-b_{R}^{2/3}\right)^{6} \right] / (2^{24}\alpha^{6}\kappa^{12}\rho^{6}q^{3}).
\end{eqnarray}
The requirement that one should recover the conventional cosmology leads to the relation $\kappa^{2}_{4}=\kappa^{4}\lambda/6b^{2/3}$. And the 4D effective cosmological constant vanishes 
when $\lambda = \frac{3}{2}\left(1-b^{1/3}\right)\frac{1}{\alpha\kappa^{2}_{4}}$. Using these 
relations, we find that the assumption $( m_{GB} > m_{RS})$  is  inconsistent with (26). \\
Here, this assumption comes from the consideration that 
the GB term is a correction to the RS gravity. However, in this case  the RS model is not recovered 
for $\alpha=0$. So the condition $(m_{GB} < m_{RS})$ might be allowed. Let us study such cases. \\
$\bullet$Case 1. $\left(m_{RS}^{8}  \hspace{0.3em}\raisebox{0.4ex}{$>$}\hspace{-0.75em}\raisebox{-.7ex}{$\sim$}\hspace{0.3em}\ 2m_{GB}^{8}\right)$ \\
Setting $\left(\kappa^{4}\alpha\rho^{2}\right)^{-1}$ and $\alpha\mu/a^{4}$ as small variables, the Friedmann 
equation is approximated as 
\begin{eqnarray}
H^{2} + \frac{k}{a^{2}}= -\frac{1}{4\alpha}+\left(\frac{\kappa^{2}\rho}{16\alpha}\right)^{2/3} - \frac{1}{12\alpha}\left(\frac{b^{3/2}}{2\alpha\kappa^{4}\rho^{2}}\right)^{2/3} \nonumber\\
\sim -\frac{1}{4\alpha}+\left(\frac{\kappa^{2}\lambda}{16\alpha}\right)^{2/3} - \frac{1}{12\alpha}\left(\frac{b^{3/2}}{2\alpha\kappa^{4}\lambda^{2}}\right)^{2/3}\nonumber\\
 + \left(\frac{2}{3\lambda}\left(\frac{\kappa^{2}\lambda}{16\alpha}\right)^{2/3} - \frac{1}{9\alpha\lambda}\left(\frac{b^{3/2}}{2\alpha\kappa^{4}\lambda^{2}}\right)^{2/3}\right)\sigma.
\end{eqnarray}
In order to obtain standard form we get the following relations:
\begin{eqnarray}
\kappa^{2}_{4} = \frac{2}{\lambda}\left(\frac{\kappa^{2}\lambda}{16\alpha}\right)^{2/3} - \frac{1}{3\alpha\lambda}\left(\frac{b^{3/2}}{2\alpha\kappa^{4}\lambda^{2}}\right)^{2/3}, \\
\kappa^{4}\lambda^{2} = \frac{1}{3\alpha B^{2/3}}
\left(16+ 4\beta + 8B^{1/3}+\beta B^{1/3}+4B^{2/3} \right. \nonumber \\
\left.+b^{3/2}(12+3B^{1/3}+B^{2/3})\right),
\end{eqnarray}
where $\beta = \sqrt{b\left(16b+9\right)}$ and $B=8+9b+3\beta$.
By Eq. (47) we find that the assumption $\left(m_{RS}^{8}  \hspace{0.3em}\raisebox{0.4ex}{$>$}\hspace{-0.75em}\raisebox{-.7ex}{$\sim$}\hspace{0.3em}\ 2m_{GB}^{8}\right)$  is consistent with Eq. (26) 
(see figure \ref{lambda}). \\
\\
$\bullet$Case 2. $\left(  1\hspace{0.3em}\raisebox{0.4ex}{$<$}\hspace{-0.75em}\raisebox{-.7ex}{$\sim$}\hspace{0.3em}m_{RS}^{8}/m_{GB}^{8}\hspace{0.3em}\raisebox{0.4ex}{$<$}\hspace{-0.75em}\raisebox{-.7ex}{$\sim$}\hspace{0.3em}2\right)$  \\
 At late times the Friedmann equation is approximated as
 \begin{eqnarray}
H^{2} + \frac{k}{a^{2}} = \frac{7b^{1/3}-9}{36\alpha} +\frac{\kappa^{4}\lambda^{2}}{36b^{2/3}}
-\frac{\alpha\kappa^4\lambda^2(\mu_{L}+\mu_{R})}{9b^{4/3}a^{4}}+\frac{7(\mu_{L}+\mu_{R})}{18b^{1/3}a^{4}}\nonumber\\
+\left(\frac{\kappa^{4}\lambda}{18b^{2/3}}-\frac{2\alpha\kappa^{4}\lambda(\mu_{L}+\mu_{R})}{9b^{4/3}a^{4}}\right)\sigma 
+\left(\frac{1}{36b^{2/3}}-\frac{\alpha(\mu_{L}+\mu_{R})}{9b^{4/3}a^{4}}\right)\kappa^4\sigma^{2}.~~~~
\end{eqnarray}
In order to obtain the standard form of the Friedmann equation we get the following relations:
\begin{eqnarray}
&&\kappa_{4}^{2} = \frac{\kappa^{4}\lambda}{6b^{2/3}}-\frac{2\alpha\kappa^{4}\lambda(\mu_{L}+\mu_{R})}{3b^{4/3}a_{*}^{4}}, \\
&&\kappa^{4}\lambda^{2} = \frac{9b^{2/3}-7b}{\alpha},
\end{eqnarray}
where $a_{*}$ is the scale factor today. From Eq. (A.20) we find that the assumption $\left(  1\hspace{0.3em}\raisebox{0.4ex}{$<$}\hspace{-0.75em}\raisebox{-.7ex}{$\sim$}\hspace{0.3em}m_{RS}^{8}/m_{GB}^{8}\hspace{0.3em}\raisebox{0.4ex}{$<$}\hspace{-0.75em}\raisebox{-.7ex}{$\sim$}\hspace{0.3em}2\right)$ is satisfied if $5/11\hspace{0.3em}\raisebox{0.4ex}{$<$}\hspace{-0.75em}\raisebox{-.7ex}{$\sim$}\hspace{0.3em}\alpha l^{-2} \hspace{0.3em}\raisebox{0.4ex}{$<$}\hspace{-0.75em}\raisebox{-.7ex}{$\sim$}\hspace{0.3em} 17/36. $ This condition is consistent with Eq.(27).

\end{document}